\documentclass[11pt,4apaper]{article}
\usepackage[latin9]{inputenc}
\usepackage[english]{babel}
\usepackage{verbatim}
\usepackage[algoruled]{algorithm2e}
\usepackage{amsmath,amssymb,array}
\usepackage{amsthm}
\usepackage[dvips]{graphicx}
\begin{document}

\newtheorem{teo}{Theorem}[section]
\newtheorem{lem}[teo]{Lemma}
\newtheorem{Pro}[teo]{Property}
\newtheorem{cor}[teo]{Corollary}
\newtheorem{Cl}[teo]{Claim}
\theoremstyle{definition}
\newtheorem{Def}{Definition}[section]
\newtheorem{PB}{Problem}
\newtheorem{example}{Example}[section]
\newcommand{\nota}[1]{{\sffamily\bfseries #1}\marginpar{\framebox{$\mathbf{\Leftarrow}$}}}
\newcommand{\grafo}{\cal G}
\newcommand{\greedy}{\textsc   Greedy-CC}
\newcommand{\Algo}{\textsc Greedy-reduced-CC}
\newcommand{\NB}[1]{{\framebox{\small\bfseries #1}}\marginpar{\framebox{$\mathbf{\Leftarrow}$}}}

\title{Approximating Clustering of Fingerprint Vectors with Missing Values}
\author{Paola Bonizzoni \and Gianluca Della Vedova \and Riccardo Dondi
}

\date{}
\maketitle

\begin{abstract}
The problem of clustering fingerprint vectors is an interesting problem in
Computational Biology that has been proposed in \cite{figueroa04:_clust_dna}.
In this paper we show some improvements in closing the gaps between the known
lower bounds and upper bounds on the approximability of some variants of the
biological problem. Namely we are able to prove that
the problem is APX-hard even when each fingerprint contains only two unknown position.
Moreover we have studied some variants of the orginal problem, and we give two
$2$-approximation algorithm for the IECMV and OECMV problems when the number of
unknown entries for each vector is at most a constant.
\end{abstract}

\section{Introduction}
High-throughput approaches for the examination of microbial communities are
becoming increasingly important, especially after the oligonucleotide
fingerprinting strategy has found wide application, allowing the identification
of thousands of cDNA clones
\cite{Drm99,DD94,DSL96,valinsky02:_oligon_finger_genes_analy_fungal_commun_compos,valinsky02:_analysis}.
After the rDNA clone
libraries are constructed, the clones are classified by individual
hybridization experiments on DNA microarrays with a series of short DNA oligonucleotides
into {\em clone types} or {\em operational taxonomic units} (OTUs), where a an
OTU is a set of DNA clones sharing the same set of oligonucleotides that have
successfully hybridized. Once
classified, the nucleotide sequence of representative clones from each OTU can
then be obtained by DNA sequencing to provide phylogenetic descriptions of the
microorganisms. One of the key features of this strategy is that after a
comprehensive database, that correlates hybridization
patterns with nucleotide sequence data, has been compiled,
little additional rDNA clone sequencing will be required,
resulting in significant reduction of cost and
effort. The effectiveness of this general strategy has been demonstrated in
the biotechnology arena, where it is currently being used to screen and
identify millions of cDNA clones \cite{Drm99}.

The oligonucleotide fingerprinting method is commonly used to study
DNA clone libraries. Such method naturally leads to a combinatorial problem
where for each oligonucleotide we are give a fingerprint over the
alphabet $\{ 0,1,N\}$, where the values $0$ or $1$ means respectively that the a
hybridization has happened or not with a certain clone, while the value $N$
stands for the fact that
we are unable to determine if the hybridization has happened or not (typically
it is due to the fact that there are two control signals, and the values between
those two control
signals mean that either result might have happened).

The combinatorial problem that naturally arises is called \textsc{CMV}. In such
problem we are given a set of fingerprints and we would like to change each
$N$-symbol in the input fingerprints into $0$ or $1$, so that the total number
of distinct fingerprints (over $\{0,1\}$) is minimized. Actually we are not
interested into the actual fingerprints over  $\{0,1\}$, but only in determining
the clusters of fingerprints.

Unfortunately the problem is NP-hard, therefore it
is important to study if some restrictions become tractable. For instance it is
possible to restrict the problem to instances where each input
fingerprint contains at most $p$ $N$-symbols, and we will call such problem
CMV($p$).
It is already known that CMV(2) is
NP-hard\cite{figueroa05:_aprox_clust_finger_vector_missin_values}, while
CMV(1) can be solved in polynomial-time\cite{figueroa04:_clust_dna}, so
for all interesting values of $p$ we have to concentrate on developing
approximation algorithms.
CMV(p) is known to be approximable within factor $2^p$ \cite{figueroa04:_clust_dna}
and
$\min (1 + \ln n, 2+p\ln l)$\cite{figueroa05:_aprox_clust_finger_vector_missin_values},
where $l$ is the lenght of the fingerprint vectors. In this paper we strengthen
the NP-hardness result proving that CMV(2) is APX-hard, that is it cannot be
approximated within an arbitrarily small $(1+\epsilon)$-factor polynomial-time
algorithm unless P=NP \cite{listproblems}.

Moreover we will study two related optimization problems, namely \textsc{IECMV} and
\textsc{OECMV}
where we want to minimize the number of pairs of compatible fingerprints that
are not clustered together and the number of pairs of incompatible fingerprints
that are clustered together, respectively.
Again we are interested in the restrictions of IECMV and OECMV with at most $p$
missing values in each fingerprint (those problems are denoted by IECMV($p$) and
OECMV($p$) respectively).  The IECMV($p$) problem is known to be approximable within factor $2^{2p-1}$
for any $p=O(\log n)$
\cite{figueroa05:_aprox_clust_finger_vector_missin_values}, while the
restriction of \textsc{OECMV} where no two compatible fingerprint vectors
can have value $N$ at the same position can be approximated within factor
$2(1-\frac{1}{2p})$\cite{figueroa05:_aprox_clust_finger_vector_missin_values}.

In this paper we improve those approximation results, proving that both
IECMV($p$) and OECMV($p$) problems are APX-hard, and we
show that a simple greedy algorithm achieves a $2$ approximation ratio for both problems.

\section{Preliminary Definitions}
\label{defi}

In this section, we introduce some basic notations and definitions
that we will need later.
A fingerprint vector (in short \emph{fingerprint}) is a vector over the alphabet
$\{ 0,1,N \}$, where
$1$ represents a hybridization, $0$ represents no hybridization
and $N$ represents unknown data (that is we are unable to determine if
hybridization has happened or not). In all instances of the problems that we
will study, all fingerprints have the same length, that is they contain the same
number of elements. Usually we will denote by $l$ the lenght of a fingerprint.

Two fingerprints vectors $v_1=\langle v_1[1], v_1[2], \ldots ,v_1[l]\rangle$ and $v_2=\langle
v_2[1],  \ldots ,v_2[l]\rangle$ are \emph{compatible} if for any
position $i$ where they differ, at least one of  $v_1[i]$ and $v_2[i]$ is equal
to $N$.
A resolved vector $r=\langle r[1],  \ldots ,r[k]\rangle$ of a fingerprint vector $v=\langle v[1],
\ldots ,v[k]\rangle$ is a vector over alphabet $\{0,1 \}$
such that for each $1\leq i\leq l$, if $v[i]\neq N$ then $v[i]=r[i]$, that is $r$ and $v$
agree on each position where $v$ is not unknown. Sometimes it is useful the
analyze the effect of a parameter, the maximum number of $N$s allowed in a
fingerprint; we will denote by $p$ such parameter. We are now ready to present
the problem we will study.

\textsc{Clustering with $p$ missing values}(CMV($p$)): We are given a
set $F$ of fingerprint vectors with at most $p$ $N$s and we want to
partition $F$ into disjoint subsets $F_1,\ldots, F_k$ such that any two vectors in
$F_i$ are compatible and the cardinality of the partition is minimized.

\textsc{Inside Clustering with $p$ missing values}(IECMV($p$)): We are given a
set $F$ of fingerprint vectors with at most $p$  $N$s and we want to
partition $F$ into disjoint subsets $F_1,\ldots, F_k$ such that any two vectors in $F_i$
are compatible and the number of compatible pairs of vectors within the same clusters
is maximized.

\textsc{Outside Clustering with $p$ missing values}(OECMV($p$)): We are given a
set $F$ of fingerprint vectors with at most $p$  $N$s and we want to
partition $F$ into disjoint subsets $F_1,\ldots, F_k$ such that any two vectors in $F_i$
are compatible and the number of compatible pairs of vectors belonging to different
clusters is minimized.

Notice that for all the aforementioned problems, the instance is a set $F$ of
fingerprints and the output is a partition  of $F$ where in a set of the
partition there are only pairwise compatible fingerprints. It is easy to notice
that pairwise compatibility is a sufficient condition to prove the existence of
a common resolution for all fingerprints in the set. For simplicity's sake in
the following we will denote by $n$ the number of fingerprints in an instance $F$.

\section{An approximation algorithm for IECMV($p$) and OECMV($p$)}

In this section we present an approximation algorithm for both IECMV($p$) and
OECMV($p$) problems, where $p$ is any fixed constant. We are able to provide two
different analysis, one for each
problem, showing that we achieve a $2$-approximation for both problems.

Given a set $F$ of fingerprints, since $p$ is a constant we are able (in
$O(2^pn)l$ time) to compute the set
$R=\{r_1, \ldots , r_k\}$ of all possible resolved fingerprints that are compatible with
at least one fingerprint in $F$.
Given a resolved fingerprint $r$, we denote by $s(r)$ the set
of fingerprints in $F$ that are compatible with $r$, and denote by
$p(s(r))$  the set of pairs of vectors in $s(r)$.
The \emph{degree} of a fingerprint $r$, denoted by $d(r)$,
is defined as the cardinality of $s(r)$.

The algorithm constructs a partition $P$ of $F$ greedily as follows: initially
let $P$ be an empty set and let $U$ be equal to $F$.
At each iteration the algorithm computes the resolved fingerprint $r$ of maximum
degree (i.e. $r$ is the resolved fingerprint compatible with the maximum number
of fingerprints in $U$),  adds $s(r)$ as a set of the solution $P$ and removes
all fingerprints in $s(r)$ from $U$. The algorithm iterates such step until $U$
is empty.

\subsection{Analysis for IECMV($p$)}

Let $S=\{s_1,\ldots ,s_k\}$ be a solution $S$ of IECMV($p$).
The value of $S$
is the number of
compatible fingerprints vectors co-clustered by $S$ and is denoted
by $V(S)$. It holds that $V(S) = \sum_{i=1}^{t} |P(s_i)|$, where $P(s_i)$ is the
set of pairs of fingerprints in $s_i$. Generalizing such notion, we denote by
$P(S)$ the set of all the pairs co-clustered in the partition
$S$, that is $P(S)=\cup_{i=1}^{|S|} P(s_i)$.
Let $W \subseteq U$ be a subset of fingerprint vectors,
we denote by $P(S, W )$ the set of pairs $(x,y)$ in $P(S)$
such that at least one of $x$, $y$ is in $W$.

In the following we will show that the algorithm has approximation factor $2$.
The algorithm computes a sequence $\langle r_1, \ldots , r_k\rangle$ of resolved fingerprints,
one at each iteration. At the $i$-th iteration the algorithm contructs a set of
the partition containing $r_i$ and all fingerprints that are compatible with
$r_i$ and have not been put in a partition during one of the previous
iterations (we will denote such set by $s_i$). For ease of the analysis, we will
denote by $U_i$ the set $U$ at the beginning of the $i$-th iteration,
consequently $U_1=F$, $U_{i+1} = U_i \setminus s_i$, for $1 \leq i < k $,
where $k$ is the number of sets in the
output partition. Recall that the partition output by the algorithm is denoted by
$S=\{s_1, \ldots , s_k\}$. The optimal partition is denoted by $Opt=\{opt_1, \ldots , opt_l\}$,
where $l$ can be different from $k$.

By definition, the value of the optimal solution is $|P(Opt)|$; our goal will be to
show that $|P(Opt)|\leq 2|P(S)|$. We introduce some sets as follows: $P(Opt,1) =
P(Opt, s_1)$, and $P(Opt,i+1) =  P(Opt, s_i) \setminus \bigcup_{1 \leq j \leq i}
P(Opt, j)$ for $1\leq i<k$. A fundamental property is that $\{ P(Opt,i) : 1\leq i<k\}$
is a
partition of $P(Opt)$.

In fact the sets $P(Opt, i)$ are  disjoint by construction. Since
$S=\{s_1,\ldots,s_k\}$ is a partition of $F$, then $P(Opt)=\bigcup P(Opt, s_i)$. Let $(x,y)$ be a
pair of $P(Opt)$. W.l.o.g. we can assume that $x\in s_i$,  $y\in s_j$, with $i\leq j$.
Then $(x,y)\in P(Opt, s_i)$ and $(x,y)$ does not belong to any $P(Opt, h)$ with
$h<i$, therefore $(x,y)\in P(Opt, i)$. Consequently the sets $P(Opt, i)$ are a
partition of $P(Opt)$, and the value of the optimal solution is equal to $\sum_i
|P(Opt, i)|$.

Consequently,
in order to prove that our greedy algorithm achieves a $2$ approximation, it
suffices to show that, for each $i$, $|P(Opt, i)|\leq 2|P(s_i)|$.

\begin{lem}
\label{lem-pres-factor}
Let $S=\{s_1, \ldots ,s_k\}$ be the solution computed by the algorithm, and let $Opt$
be an optimal solution. Then $|P(Opt, i)|\leq 2|P(s_i)|$ for $1\leq i\leq k$.
\end{lem}

\begin{proof}
Let  $s_i$ be  the set added to the solution $S$ at the $i$-th
step of the algorithm.
All pairs in $P(Opt, i)$ must belong to $U_i× U_i$, by definition of $P(Opt,
i)$. Each element $x$ in $U_i$ is in the same set of the optimal solution with
at most $|s_i|-1$ other
fingerprints of $U_i$, for otherwise the algorithm would not have chosen $s_i$ at
the $i$-th iteration, but $x$ and all fingerprints in $U_i$ that are in the same
set of the optimal solution.
By definition of $P(Opt, i)$, there are at most $|s_i|(|s_i|-1)$ pairs in
$P(Opt, i)$, which completes the proof, since in $s_i$ there are
$|s_i|(|s_i|-1)/2$ pairs.
\end{proof}

It is easy to see that approximation factor is tight.
Consider three resolved vectors $r_1$, $r_2$, $r_3$ and four fingerprint
vectors $\{ f_1, f_2, f_3, f_4  \}$
such that $s(r_1)=\{ f_1, f_2 \}$, $s(r_2)=\{ f_1, f_3 \}$,
$s(r_2)=\{ f_2, f_4 \}$. The approximation algorithm choose
$s(r_1)$ as the first set and then $\{f_3\}$, $\{f_4\}$ as the sets to complete
the partition. Thus value of the approximated solution is $1$, since
one pair is selected.
It is easy to see that the optimal solution consists of set
$s(r_2)=\{ f_1, f_3 \}$, $s(r_2)=\{ f_2, f_4 \}$, thus having value
$2$.

\subsection{Analysis for OECMV($p$)}

The analysis in this section follows the one for IECMV($p$).
Let $S=\{s_1,\ldots ,s_k\}$ be a solution $S$ of OECMV($p$).
The value of $S$ is the number of
compatible fingerprints vectors that are not co-clustered in $S$ and is denoted
by $V(S)$. It holds that $V(S) = \frac{1}{2}\sum_{i=1}^{k} |L(s_i)|$, where $L(s_i)$ is the
set of pairs $(x,y)$ of compatible fingerprints where exactly one of $x$ and $y$ is in
$s_i$. Generalizing such notion, we denote by $L(S)$ the set of all unordered
pairs of compatible fingerprints
that are not co-clustered in the partition
$S$, that is $L(S)=\cup_{i=1}^{|S|} L(s_i)$. Notice also that each pair in $L(S)$
appears in exactly two sets $L(s_i)$, therefore $|L(S)|=\frac{1}{2}\sum_{i=1}^{|S|} |L(s_i)|$.
Let $W \subseteq U$ be a subset of fingerprint vectors,
we denote by $L(S, W )$ the set of pairs $(x,y)$ in $L(S)$
such that at least one of $x$, $y$ is in $W$.

In the following we will show that the algorithm has approximation factor $2$.
The algorithm computes a sequence $\langle r_1, \ldots , r_k\rangle$ of resolved fingerprints,
one at each iteration. At the $i$-th iteration the algorithm contructs a set of
the partition containing $r_i$ and all fingerprints that are compatible with
$r_i$ and have not been put in a partition during one of the previous
iterations (we will denote such set by $s_i$). For ease of the analysis, we will
denote by $U_i$ the set $U$ at the beginning of the $i$-th iteration,
consequently $U_1=F$, $U_{i+1} = U_i \setminus s_i$, for $1 \leq i < k $,
where $k$ is the number of sets in the
output partition. Recall that the partition output by the algorithm is denoted by
$S=\{s_1, \ldots , s_k\}$. The optimal partition is denoted by $Opt=\{opt_1, \ldots , opt_l\}$,
where $l$ can be different from $k$.

By definition, the value of the optimal solution is $|L(Opt)|$; our goal will be to
show that $2|L(Opt)|\geq |L(S)|$. We introduce some sets as follows: $L(Opt,1) =
L(Opt, s_1)$, and $L(Opt,i) =  L(Opt, s_i) \setminus \bigcup_{1 \leq j < i}L(Opt, j)$ for
$1\leq i \leq k$. A fundamental property is that $\{ L(Opt,i) : 1\leq i \leq k\}$ is a
partition of $L(Opt)$.
In fact the sets $L(Opt, i)$ are  disjoint by construction. Since
$S=\{s_1,\ldots,s_k\}$ is a partition of $F$,
then $L(Opt)=\bigcup L(Opt, s_i)$. Let $(x,y)$ be a
pair of $L(Opt)$. W.l.o.g. we can assume that $x\in s_i$,  $y\in s_j$, with $i\leq j$.
Then $(x,y)\in L(Opt, s_i)$ and $(x,y)$ does not belong to any $L(Opt, h)$ with
$h<i$, therefore $(x,y)\in L(Opt, i)$. Consequently the sets $L(Opt, i)$ are a
partition of $L(Opt)$, and the value of the optimal solution is equal to
$\sum_i |L(Opt, i)|$.

Similarly we introduce the sets $L(S,1)=L(s_1)$,
$L(S,i) =  L( s_i) \setminus \bigcup_{1 \leq j < i}L(S, j)$ for $1 \leq i \leq k$.
A fundamental property is that $\{ L(S,i) : 1\leq i \leq k\}$ is a
partition of $L(S)$ and thus $|L(S)| = \sum_i |L(S,i)|$.
Consequently,
in order to prove that our greedy algorithm achieves a $2$ approximation, it
suffices to show that, for each $i$, $2|L(Opt, i)|\geq |L(S,i)|$.

\begin{lem}
\label{lem-pres-factor2}
Let $S=\{s_1, \ldots ,s_k\}$ be the solution computed by the algorithm, and let $Opt$
be an optimal solution. Then $2|L(Opt, i)|\geq |L(S,i)|$ for $1\leq i\leq k$.
\end{lem}

\begin{proof}
Let $s_i$ be the set added to the solution $S$, at the $i$-th
step of the algorithm. Given a fingerprint $x \in s_i$, we define $C(x)$ as the set
of all fingerprints in $U_i$ that are compatible with $x$, and $D(x)$ as the set
$C(x)\cap L(Opt,i)$, that is the pairs in $C(x)$ that are not co-clustered in $Opt$.
Since $x$ is clustered with $|s_i| -1$ elements of $U_i$ in $S$,
there are exactly $|C(x)|-|s_i|+1$ pairs in $L(S,i)$ containing
$x$.  It follows that
$|L(S,i)| = \sum_{x \in s_i} \left( |C(x)|-|s_i|+1 \right)$.

All pairs in $L(Opt, i)$ must belong to $(U_i\setminus  U_{i+1}) × U_i$, by definition
of $L(Opt,i)$ (for simplicity, we will assume that $U_{k+1}=\emptyset$).
Notice that, by construction of $s_i$, $|D(x)| \geq |C(x)|-|s_i|+1$.
Clearly $L(Opt,i) = \cup_{x\in U_i}D(x)$, by definition of $D(x)$. Since each pair
$(y,z)\in L(Opt,i)$ appears only in the two (not necessarily distinct) sets $D(y)$
and $D(z)$, then $|L(Opt,i)| \geq\frac{1}{2} \sum_{x\in U_i}|D(x)| \geq
\frac{1}{2}\sum_{x \in U_i} \left( |C(x)|-|s_i|+1 \right) $
and the proof is completed.
\end{proof}

It is easy to see that also in this case the approximation factor is tight.
Consider three resolved vectors $r_1$, $r_2$, $r_3$ and four fingerprint
vectors $\{ f_1, f_2, f_3, f_4  \}$
such that $s(r_1)=\{ f_1, f_2 \}$, $s(r_2)=\{ f_1, f_3 \}$,
$s(r_2)=\{ f_2, f_4 \}$. The approximation algorithm choose
$s(r_1)$ as the first set and then $\{f_3\}$, $\{f_4\}$ as the sets to complete
the partition. Thus the value of the approximated solution is $2$, since
the pairs of compatible fingerprint vectors
that are not co-clustered  are $(f_1, f_3)$ and $(f_2, f_4)$.
It is easy to see that the optimal solution consists of set
$s(r_2)=\{ f_1, f_3 \}$, $s(r_2)=\{ f_2, f_4 \}$, hence the only
pair of compatible fingerprint vectors not co-clustered in the optimal solution
is $(f_1, f_2)$ and the cost of the optimal solution is $1$.

\section{APX-hardness of CMV($2$)}

In this section we will prove that CMV($p$) is APX-hard via an $l$-reduction
from minimum vertex cover on cubic graph, whose APX-hardness has been proved in \cite{AK92}.

In particular we will combine two l-reductions: $(1)$ from
minimum  vertex cover on a graph $G$ to
minimum  vertex cover on a gadget graph;
$(2)$ from minimum vertex cover on a gadget graph to CMV($p$).

\subsubsection*{First Reduction}

First we will define gadget graphs.
Given a graph $G=(V,E)$ for each vertex $v_i \in V$ we define a vertex gadget $VG_i$
consisting of $5$ vertex. Three vertices,
$c_{i_1}$, $c_{i_4}$, $c_{i_5}$ are called docking vertices.
Observe that the minimum vertex cover of a vertex gadget consists of
$2$ vertices,
$c_{i_2}$, $c_{i_3}$, and denote this cover as \emph{type 1}.
Observe that
there is a cover of $VG_i$ consisting of $3$ vertices
$c_{i_1}$, $c_{i_4}$, $c_{i_5}$, and denote this cover as \emph{type 2}.

\begin{figure}
\begin{center}
\includegraphics[width=4cm]{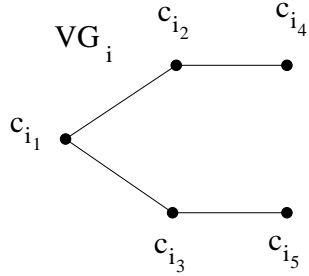}
\caption{A vertex gadget $VG_i$} \label{vertex-gadget}
\end{center}
\end{figure}

For each edge $(v_i,v_j)$ we define an edge gadget $EG_{i,j}$
joining vertex gadgets $VG_i$, $VG_j$ in two of their docking vertices.
An edge gadget consists of six vertices, the two docking vertices
shared with the vertex gadgets and other four vertices.

\begin{figure}
\begin{center}
\includegraphics[width=12cm]{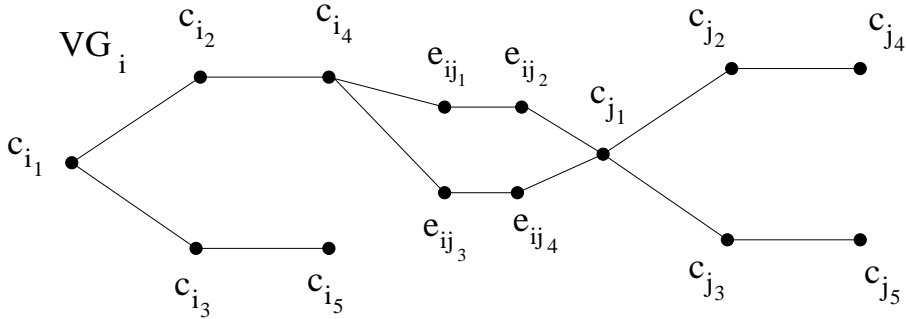}
\caption{An edge gadget $EG_{ij}$} \label{edge-gadget}
\end{center}
\end{figure}

\begin{teo}
Let $C \subseteq V$ be a cover of $G$, with $|C|=k$. Then there is a cover
of the graph gadget of size $3k+2(n-k)+2m$.
\end{teo}
\begin{proof}
Consider a vertex $v_i$ in $C$, associate with the corresponding vertex gadget $VG_i$
a cover of type 2 (of size $3$). For each vertex $v_j \notin  C$, associate with the corresponding vertex gadget $VG_j$ a cover of type 1 (size $2$).
Observe that for each edge gadget at least one of the adjacent vertex gadget
has a type 2 cover. Thus we just need to cover two vertices for each
edge gadget to obtain a cover of each edge gadget and thus of the entire graph gadget.
\end{proof}

\begin{lem}
Let $C$ be a cover of the graph gadget of size
$3k+2(n-k)+2m$. Then we can compute in polynomial time a vertex cover
of size at most $3k+2(n-k)+2m$ such that it has only
cover type 1 and type 2 and such that for each pair of adjacent
vertex gadgets at least one has a cover of type 2.
\end{lem}
\begin{proof}
It is easy to see that if a vertex gadget has not a cover of type 1
we can substitute this solution with a cover of type 2, obtaining
a solution with at least the same size.
Now assume that two  adjacent vertex gadgets $VG_i$, $VG_j$ have both a cover of type 1.
Then observe that the edge gadget $EG_{ij}$ must be cover with at least $4$
vertices. By covering $VG_j$ with a cover of type 2, the edge gadget $EG_{i,j}$
needs to be covered just with $2$ elements thus obtaining a cover of size less than
$3k+2(n-k)+2m$.
\end{proof}

\begin{teo}
Let $C $ be a cover of the graph gadget of size $3k+2(n-k)+2m$.
Then there is a cover of the graph $G$ of size $k$.
\end{teo}
\begin{proof}
Consider a vertex cover of size $3k+2(n-k)+2m$.
Then from previous lemma we can construct a solution of size at most
$3k+2(n-k)+2m$ and such that for each edge gadget $EG_{ij}$ at least one of
$VG_i$, $VG_j$ is of type 2.
Thus, we can define a cover for $G$ taking all the vertices
corresponding to vertex gadgets with a cover of type 2.
Since there are at most $k$ vertex of this kind the theorem follows.
\end{proof}

Since a vertex cover of a cubic graph contains at least $|V|/4$ vertices and
$|E|= \frac{3}{2}|V|$, it follows that the above reduction is an l-reduction.

\subsubsection*{Second Reduction}

Now we reduce minimum vertex cover on gadget graph to CMV($p$). The idea in our
description is that it is possible to assign a resolved fingerprint to each
vertex and an unresolved fingerprint to each edge.
The set of unresolved fingerprints will be our instance of CMV($p$), and all
interesting solutions will pick their resolved fingerprints from those assigned
to the vertices. More precisely we will show that each unresolved fingerprint
(assigned to an edge)
will be resolved to the fingerprints assigned to one of the endpoints of such edge.

Let $n$ denote the number of vertex gadgets. Each fingerprint consists of $n$
chunks of $7$ positions, and each vertex in the vertex gadget $VG_i$ consists
only of $0$s, except for the $i$-th chunk.
Denote with $v(c_{i_1})$, $v(c_{i_2})$, $v(c_{i_3})$, $v(c_{i_4})$ and $v(c_{i_5})$
the resolved vectors associated with the vertices of $VG_i$, define the $i$-th chunk
of these vectors as follows:
$v(c_{i_1})\to 1110000$, $v(c_{i_2})\to 1111100$,
$v(c_{i_3})\to 1110011$, $v(c_{i_4})\to 1001100$, $v(c_{i_5})\to 1000011$. For
example, the vertex $v(c_{i_4})$ of the $i$-th vertex gadget has fingerprint
$0^{7(i-1)} 1001100 0^{7(n-i+1)}$.

The vertices belonging exclusively to an edge gadget will have two chunks that
are not completely made of $0$s. More precisely, let $VG_i$ and $VG_j$ be two
adjacent vertex gadgets,
we denote with
$v(e_{i,j,1})$, $v(e_{i,j,2})$, $v(e_{i,j,3})$,
$v(e_{i,j,4})$ the resolved vector associated with
the vertices of the edge
gadgets $VG_{ij}$.
Only the $i$-th and the $j$-th chunks are not completely consisting of $0$s, and
those chunks are represented in Table~\ref{tab:edge-chunks1}.

Assume that
$e_{i,j,1}$, $e_{i,j,3}$ are adjacent to a vertex of $VG_i$, $c_{i_x}$, and that
$e_{i,j,2}$, $e_{i,j,4}$ are adjacent to a vertex of $VG_j$, $c_{j_y}$.
We define these resolved vectors as follows:

\begin{table}[htb]
  \centering
  \begin{tabular}{llllllll}
    \hline
    chunk &$VG_i$&$VG_j$&\hspace{1ex}&$v(e_{i,j,1})$&$v(e_{i,j,2})$&$v(e_{i,j,3})$&$v(e_{i,j,4})$\\\hline
$i$-th&$1110000$&$0000000$&&$1100000$&$0100000$&$1010000$&$0010000$\\
$j$-th&$0000000$&$1110000$&&$0100000$&$1100000$&$0010000$&$1010000$\\\hline
$i$-th&$1110000$&$0000000$&&$1100000$&$0100000$&$1010000$&$0010000$\\
$j$-th&$0000000$&$1001100$&&$0001000$&$1001000$&$0000100$&$1000100$\\\hline
$i$-th&$1110000$&$0000000$&&$1100000$&$0100000$&$1010000$&$0010000$\\
$j$-th&$0000000$&$1000011$&&$0000010$&$1000010$&$0000001$&$1000001$\\\hline
$i$-th&$1001100$&$0000000$&&$1001000$&$0001000$&$1000100$&$0000100$\\
$j$-th&$0000000$&$1110000$&&$0100000$&$1100000$&$0010000$&$1010000$\\\hline
$i$-th&$1001100$&$0000000$&&$1001000$&$0001000$&$1000100$&$0000100$\\
$j$-th&$0000000$&$1001100$&&$0001000$&$1001000$&$0000100$&$1000100$\\\hline
$i$-th&$1001100$&$0000000$&&$1001000$&$0001000$&$1000100$&$0000100$\\
$j$-th&$0000000$&$1000011$&&$0000010$&$1000010$&$0000001$&$1000001$\\\hline
$i$-th&$1000011$&$0000000$&&$1000010$&$0000010$&$1000001$&$0000001$\\
$j$-th&$0000000$&$1110000$&&$0100000$&$1100000$&$0010000$&$1010000$\\\hline
$i$-th&$1000011$&$0000000$&&$1000010$&$0000010$&$1000001$&$0000001$\\
$j$-th&$0000000$&$1001100$&&$0001000$&$1001000$&$0000100$&$1000100$\\\hline

$i$-th&$1000011$&$0000000$&&$1000001$&$0000010$&$1000001$&$0000001$\\
$j$-th&$0000000$&$1000011$&&$0000010$&$1000010$&$0000001$&$1000001$\\\hline
  \end{tabular}
  \caption{Possible values of fingerprints for an edge gadget}
  \label{tab:edge-chunks1}
\end{table}

Next we discuss the properties of the resolved vectors defined above.
Each pair of resolved vectors associated with adjacent vertices has hamming distance $2$.
Each pair of resolved vectors associated with not adjacent vertices has
hamming distance at least $3$.

Now we construct the instance of the problem, that is
the fingerprint vectors. We associate a fingerprint with each
edge of the graph gadget. Now, let $y=(a,b)$ be an edge of the graph gadget,
$v_a$ and $v_b$ the resolved vectors associated with vertices $a$ and $b$ respectively,
we associate with $y$ the fingerprint vector $v_y$ as follows:
for each position $l$ such that $v_a[l] = v_b[l] $, it follows $v_e[l]:=v_a[l]$;
for each position $l$ such that $v_a[l] \neq v_b[l] $, it follows $v_e[l]:=N$.

\begin{lem}
Each fingerprint vector has exactly two positions with value $N$.
\end{lem}
\begin{proof}
It is easy to see that by construction two resolved vectors
associated with an edge differ in exactly two positions. Thus, the fingerprint vector
associated with that edge has value $N$ in those positions.
\end{proof}

A fundamental property of the instance of CMV($p$) is the following:

\begin{lem}
Two fingerprint vectors can have a common resolution only it the edges encoded
by such fingerprints share a common vertex.
\end{lem}
\begin{proof}
First observe that by construction each fingerprint vector $f_i$
can have at most $4$ resolutions. Moreover, if $r_{i_1}$ and
$r_{i_2}$ are resolutions of $f_i$ having hamming distance $2$,
any other resolution have hamming distance $1$ from both $r_{i_1}$
and $r_{i_2}$.
Let $f_i$ be a fingerprint vector encoding edge
$e_i=(i_1, i_2)$ and let $f_j$ be a fingerprint vector encoding edge
$e_j=(j_1, j_2)$. 
There is at least one pair of resolved vectors associated with
the endpoints of $e_i$ and $e_j$ having hamming distance at least $3$;
assume w.l.o.g. those vectors are $r(i_1)$ and $r(j_1)$.
Note that none of  $r(i_1)$ and $r(j_1)$ can be a
common resolution for both $f_i$ and $f_j$.
Any resolution $r_i^*$ of $f_i$ different from $r(i_1)$ and $r(i_2)$,
has hamming distance $1$ from $r(i_1)$. Similarly,
any resolution $r_j^*$ of $f_j$ different from
$r(j_1)$ and $r(j_2)$
has hamming distance $1$ from $r(j_1)$.
Thus, $r_i^*$ and $r_j^*$ have hamming distance at least $1$ and thus are not be identical.
It follows that none of $r_i^*$ and $r_j^*$ can be a
common resolution for both $f_i$ and $f_j$.
Thus $f_i$ and $f_j$ have a common resolution only if
$r(i_2)$ and $r(j_2)$ are the same vector, that is they encode the same vertex.
\end{proof}

\begin{teo}
Let $C $ be a cover of the graph gadget of size $3k+2(n-k)+2m$.
Then there is a solution of CMV($p$) of size $3k+2(n-k)+2m$.
\end{teo}
\begin{proof}
Consider a vertex cover of size $3k+2(n-k)+2m$.
Thus, we can define a solution of CMV($p$) taking as resolution
the set of vertices associated with the cover.
\end{proof}

\begin{teo}
Let $C $ be a solution of CMV($p$) of size $3k+2(n-k)+2m$, then there is a
cover of the graph gadget of size $3k+2(n-k)+2m$.
\end{teo}
\begin{proof}
Consider a solution for CMV($p$). If a fingerprint vector is associated with
a resolved vector not associated with a vertex of the gadget graph,
then this resolution is not common to any other fingerprint vector of the instance.
Thus, we can replace it with a resolved vector associated with a vertex of the graph without
increasing the size of the solution.
Then for each resolution chosen, add the corresponding vertex to
the cover of the gadget graph.
\end{proof}

It is easy to see that also this second reduction is an l-reduction.

\section{MAX-SNP hardness of IECMV($2$)}

In the following section we prove that IECMV($p$) is MAX-SNP hard
via an l-reduction from Maximum Independent Set on Cubic Graphs (MISCG).
Let $G=(V,E)$ be a cubic graph, the MISCG problem asks for the
subset $V' \subseteq V$ of maximum cardinality, such
that vertices in $V'$ are not adjacent.

We associate with a vertex $v_i$ of $V$ a set of $9$
fingerprint vectors. First we introduce a set of $8$ resolved vectors,
$C_i=\{ c_{i_1},c_{i_2},c_{i_3},c_{i_4},c_{i_5},c_{i_6},c_{i_7}, c_{i_8} \}$,
such that the resolved vectors in $C_i$ are possible solutions of the fingerprint vectors.
We represent this situation through a graph, denoted as
\emph{compatibility graph} $CG_i$, such that the resolved
vectors in $C_i$ are the vertices of $CG_i$, while the fingerprint vectors
are the edges of $CG_i$. A fingerprint vector associated with an edge
$(c_{i_u},c_{i_v})$ can be resolved by both $c_{i_u}$ and $c_{i_v}$
and by no other resolved vector in $C=\bigcup_i C_i$.
Three vertices of $CG_i$, $c_{i_1}$, $c_{i_3}$ and $c_{i_8}$
are called docking vertices.

For each edge $e=(v_i,v_j) \in E$, define a fingerprint vector that is compatible with
a resolved vector associated with a docking vertex of $CG_i$ and
a resolved vector associated with a docking vertex of $CG_j$.
We represent this fingerprint vector in the graph as an
edge, $E_{i,j}$ that joins the compatibility graphs associated with vertices  $CG_i$ and
$CG_j$.
The graph obtained will be denoted as $CG$.

Assume that $|V|=n$ and $|E|=m$.
The complete vectors of the instance of IECMV($p$) have length $5n$,
$5$ positions are associated with each vertex.
Assume w.l.o.g. that vertex $v_i$ is adjacent to vertices $v_j$, $v_h$ and $v_k$
and in particular that $c_{i_1}$ is adjacent to $CG_j$,
$c_{i_3}$ is adjacent to $CG_h$ and $c_{i_8}$ is adjacent to $CG_k$.
Complete vectors associated with
$CG_i$ are defined as follows:

\begin{itemize}

\item $c_{i_1}$ has value $1$ in the position $5j-4$,
      $c_{i_3}$ has value $1$ in the position $5h-4$,
      $c_{i_8}$ has value $1$ in the position $5k-4$.

\item for any other position not in $[5i-4,5i]$ all the complete vectors
associated with $CG_i$ have value $0$.

\item for the positions in $[5i-4,5i]$,
$c_{i_1}=11000$,
$c_{i_2}=11010$,
$c_{i_3}=10010$,
$c_{i_4}=11100$,
$c_{i_5}=10110$,
$c_{i_6}=11110$,
$c_{i_7}=11011$,
$c_{i_8}=10100$.

\end{itemize}

Ler $R$ be the set of the resolved vectors associated with vertices of the graph.
Now we construct the instance of the problem, that is
the fingerprint vectors. We associate a fingerprint vector with each
edge of the graph gadget.
For an edge of the compatibility graph, let $y=(a,b)$ be an edge of the graph gadget,
$v_a$ and $v_b$ the resolved vectors associated with $a$ and $b$ respectively,
we associate with $y$ the fingerprint vector $v_y$ as follows:
for each position $l$ such that $v_a[l] = v_b[l] $, it follows $v_y[l]:=v_a[l]$;
for each position $l$ such that $v_a[l] \neq v_b[l] $, it follows $v_y[l]:=N$.

It is easy to see that each fingerprint vector will have at most $2$
positions having value $N$, since two resolved vectors associated with
adjacent vertices will have at most hamming distance equal to $2$.

\begin{figure}
\begin{center}
\includegraphics[width=6cm]{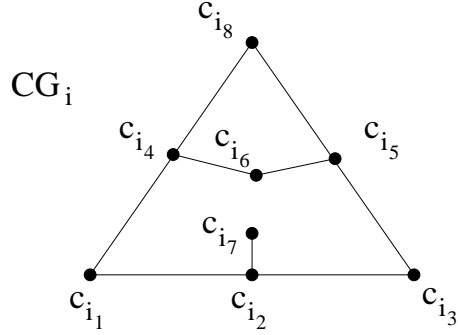}
\caption{A compatibility graph $CG_i$} \label{v-gadget-IECMV}
\end{center}
\end{figure}

\begin{figure}
\begin{center}
\includegraphics[width=14cm]{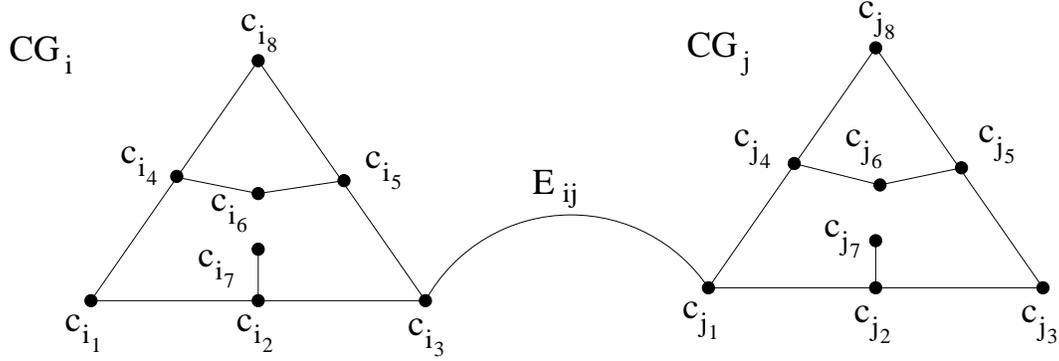}
\caption{A compatibility graph $E_{ij}$} \label{e-gadget-IECMV}
\end{center}
\end{figure}

\begin{lem}
Let $S$ be a solution of IECMV($p$), then there is a solution $S'$ having
at most the same cost and such that each resolved vector of the solution is a resolved vector
in $R$.
\end{lem}
\begin{proof}
Let $f_x$, $f_y$ be two fingerprint vectors, they are compatible if and only if
are associated with two edges incident on a common vertex.
Moreover, observe that there exists a unique resolved vertex that can be a common resolution
of both $f_x$ and $f_y$, unless they are associated with an edge incident
on the same docking vertex $c_z$. In this case they can have two common
resolutions, $r_{z_1}$ and $r_{z_2}$.
Assume that $r_{z_1}$ is associated with $C_z$, there is a single position $l$
not in $[5z-4, 5z]$ where $r_{z_1}$ has value $1$.
$r_{z_2}$ is the resolved vector having a $0$ in position $l$
and equal to $r_{z_1}$ in any other position.
Since no other vertices is compatible with $r_{z_2}$ it follows that we can
substitute $r_{z_2}$ with $r_{z_1}$ without decreasing the cost of the solution.
\end{proof}

Thus we can restrict to the solution where each set $s_v$ corresponds to a
resolved vector $r_v$ associated with a
vertex $v$ of the graph $CG$ and the fingerprint vectors associated with
(some) edges incident on $v$ are assigned to $s_v$.
In what follows we show that
for a solution of IECMV($p$) of a compatibility graph $CG_i$
we can restrict to the following cases:
\begin{itemize}
\item Solution $A$: $9$ pairs of fingerprint vectors are co-clustered; this means that
$c_{i_2}$, $c_{i_4}$ and $c_{i_5}$ are resolved vectors of the solution.

\item Solution $B$: $4$ pairs of fingerprint vectors are co-clustered; this means
that $c_{i_1}$, $c_{i_3}$, $c_{i_6}$ and $c_{i_8}$
are resolved vectors of the solution.
\end{itemize}

\begin{lem}
\label{Sol-B}
Solution $B$ is the maximum solution that has $1$ pair for each of the docking vertex of $CG_i$.
\end{lem}
\begin{proof}
Let $Z$ be a solution such that the sets associated with resolved vectors
$c_{i_1}$, $c_{i_3}$, $c_{i_8}$ have all one pair.
It is easy to see that the set associated with
resolved vector $c_{i_6}$ is the only set that can have more than one
element.
Thus the lemma follows.
\end{proof}

Let $Z$ be a solution of IECMV($p$) for $CG_i$
such that it has one set $s_x$ associated with a
resolved vector $x$ of a docking vertex.
The set $s_x$ will contain two fingerprint vectors.
If we assign the fingerprint vector associated with the edge $E_{i,j}$ incident
on $x$
to $s_x$, we gain $2$ pairs.
If we have a solution $A$ for a compatibility graph $VG_i$
and we assign the fingerprint vectors of $EG_{ij}$
to $c_{i_1}$, we gain $0$ pairs.
Note that if two adjacent compatibility graphs have as solutions
the sets corresponding to the two docking vertices, it follows
that only one of these sets can gain pairs.
Next we show that, gaining pairs from $E_{i,j}$, no solution
different from solution $B$ can become better than solution $A$.
Let $Z$ be a solution of IECMV($p$) different from solution $A$ and solution $B$.
If exactly one of the sets of $Z$ corresponds to a docking vertex, it follows that it can have at most $6$ pairs. In fact, the optimal solution in this case has one set with $3$ pairs and
three sets each one with one pair.
If exactly two of the sets of $Z$ correspond to docking vertices, it follows that it can have at most $4$ pairs. In fact, the optimal solution in this case has four sets, each one
with one pair.
Since no other solution can gain pairs from $EG_{ij}$ it follows that no solution
except solution $B$ can become better than solution $A$.

Thus the optimal solution for $CG_i$ and $EG_{i,j}$, $EG_{i,h}$,
$EG_{i,k }$ is to have solution $B$ for $CG_i$ and add fingerprint vectors
associated with $EG_{i,j}$, $EG_{i,h}$, $EG_{i,k }$ to the
sets corresponding to the docking vertices.
Each of these sets will have three elements, thus $3$ pairs, and the solution has $10$ pairs.
In what follows we will denote such a solution with solution $B$.
Moreover, any solution different from the solution constructed above, it is
worse than solution $A$.
It follows that the problem of maximizing
the number of co-clustered pairs of fingerprint
vectors consists of building an independent set of compatibility graphs (each one is associated
with solution $B$).

\begin{lem}
There exists an independent set of size $k$ if and only if exists a
solution of IECMV($p$)  having at least $10k+9(n-k)$ pairs.
\end{lem}
\begin{proof}
Let $V'$ an independent set of $G$ such that $|V'|=k$, construct a solution $S$ of
IECMV($p$) such that the component graphs associated with vertices in $V'$ have a solution of
type $B$ and any other component graph has a solution of type $A$. Then it follows
$c(S)=10k+9(n-k)$.

Now let $S$ be a solution with cost $10k+9(n-k)$. Now we can construct a solution
having at least the same cost defining for each component graph that has a cost less than $10$
a type $A$ solution.
Since the component graphs having cost $10$ must not be adjacent,
at least $k$ independent component graph must have type $B$ solution in $S$
and thus the corresponding vertices are an independent set of size $k$.
\end{proof}

Since for each cubic graph $|E|= \frac{3}{2}|V|$ and
there exists an independent set of size at least $|V|/4$,
it follows that the above reduction is an l-reduction.

\subsection{MAX-SNP hardness of OECMV($2$)}

It is easy to see that the l-reduction described above to prove the
MAX-SNP hardness of IECMV($p$) can be used also to prove the
MAX-SNP hardness of OECMV($p$).
Note that considering the set of fingerprint vectors associated with
a component graph $CG_i$ and with edges $EG_{i,j}$,
$EG_{i,h}$, $EG_{i,k}$, we can have $19$ compatible pairs of fingerprint vectors.
As in the previous reduction, the best solution for this set of fingerprint
vectors is type $B$ solution. Since type $B$ solution co-clusters
$10$ pairs of compatible fingerprint vectors, it follows that it does not co-cluster
$19-10=9$ pairs of compatible fingerprint vectors.
Similarly type $A$ solution does not co-cluster $19-9=10$ pairs of compatible vectors
and no other solution different from type $B$ solution is better
than type $A$ solution.
Hence the l-reduction for OECMV($p$) follows directly from the l-reduction
for IECMV($p$).

\bibliographystyle{abbrv}
\bibliography{abbreviations,biology,complexity,clustering,books}

\end{document}